\documentclass[]{article}
\usepackage{amsmath}
\usepackage[utf8]{inputenc}
\usepackage{graphics}
\usepackage{graphicx}
\usepackage[margin=0.8in]{geometry}
\usepackage{float}
\usepackage[linktoc=all]{hyperref}
\usepackage{cite}

\title{Barrow Holographic Dark Energy Model with GO Cut-off  - An Alternative Perspective}
\author{Nandhida Krishnan.P \footnote{Email-\it{nandhidakrishnan@cusat.ac.in}}, Titus K Mathew\footnote{Email-\it{titus@cusat.ac.in}}
	\\{*$^{\dagger}$ Department of Physics, Cochin University of Science and Technology(CUSAT)},\\ {Kochi 682-022, India.}\\{$^{\dagger}$Inter University Center for the studies on Kerala Legacy Astronomy and Mathematics, CUSAT}\\{$^{\dagger}$Center for Particle Physics, CUSAT}}
\begin{document}
	\date{}
	\maketitle

\begin{abstract}
Recently, Barrow holographic dark energy (BHDE), based on Barrow entropy, has been
proposed to describe the late acceleration of the universe. Contrary to the earlier analysis
of this model in the literature, we consider the BHDE with the Granda–Oliveros length
as IR cut-off, as a dynamical vacuum, having a constant equation of state $\omega_{\Lambda}=-1.$
We have analytically solved for the Hubble parameter and studied the evolution of
cosmological parameters. The model is compared with the observational data on Hubble
parameter (OHD36) and Supernovae type Ia (SN Ia), the pantheon data. In the absence of interaction between the dark sectors, we found that the model predicts a $\Lambda$CDM
like evolution of the universe with an effective cosmological constant. In this case, the model is found to satisfy the generalized second law (GSL), irrespective of the value of
the Barrow index. The interaction also shows the safe validity of GSL, for the extracted
value of the Barrow index, $\Delta=0.063\pm 0.029$. The thermodynamic analysis of the model
predicts an end de Sitter phase of maximum entropy. We performed a dynamical system
analysis, which reveals that the end de Sitter phase is stable. Furthermore, we performed the Information Criterion analysis using Akaike and Bayesian Information Criterion to
compare the statistical compatibility of the present model with the standard $\Lambda$CDM
model.
		
\end{abstract}




\section{Introduction}\label{intro}

Observational results from the type Ia Supernovae (SN Ia) for different redshift lead to one of the most pivotal discoveries in the twentieth century that the current universe is not just expanding but follows an accelerating expansion  \cite{SupernovaSearchTeam:1998fmf,SupernovaCosmologyProject:1998vns,WMAP:2003elm,SDSS:2003eyi,SDSS:2005xqv,WMAP:2006bqn,Frieman_2008,Planck:2015fie}. This implies that the universe, in its late epoch, can be dominated by an exotic component with negative pressure termed as dark energy, which is responsible for the current accelerated expansion. The best-fitting model for the present accelerating universe is the $\Lambda$CDM, referred to as the standard model of cosmology, which adopts the cosmological constant, $\Lambda$ as dark energy.  Despite the success in explaining the observational data \cite{Planck:2018vyg}, the $\Lambda$CDM is plagued with two major flaws. The first one is the cosmological constant problem, which arises due to the huge inconsistency between the observed value of the cosmological constant and its theoretically predicted value. The latter is the coincidence problem, the intriguing coincidence of the present value of dark energy density with that of matter, even though both have evolved differently. To alleviate these problems, one must consider the dark energy density as varying or adopt suitably modified gravity theories. The first one motivates the proposal of various dynamical dark energy models. Among these, holographic dark energy models are potentially important to interpret the dynamical dark energy \cite{WANG20171,LI20041,Huang:2004wt,Horvat:2004vn,Huang:2004ai,Feng:2007wn,Li:2009bn,PhysRevD.76.023502,Li:2013dha,Setare_2008,Wang_2008,SETARE2008111,Wang:2005ph,Setare:2007at,Sheykhi:2009zv,Feng:2007wn,Sheykhi:2009dz,Praseetha:2014deq,Praseetha:2015bjf,Pavon:2005yx,Feng:2016djj,Xu_2009,Horava:2000tb,PhysRevD.72.043524}.

The holographic dark energy models are based on the holographic principle, \cite{Susskind_1995,Bousso:2002ju,fischler1998holography} which state that the black hole entropy scales non-extensively, \cite{tHooft:1993dmi,Hooft_2001,PhysRevD.49.1912,PhysRevD.7.2333} in proportion to the area of their horizon. However, in the effective quantum field theory confined to a box, the entropy behaves extensively as, $ S \sim L^{3}\Lambda^{3}$ where $L$ is the size of the box with UV cut-off $\Lambda$. To resolve the breakdown of quantum field theory to describe a black hole, a longer distance IR and short distance UV cut-off relation was proposed by cohen et al. \cite{PhysRevLett.82.4971,myung2005holographic,2009cai} as, $L^{3}\Lambda^{4} \leq LM_{p}^{2},$ Where $\Lambda^4$ is the maximum energy density in the effective theory and $M_{p}$ is the Planck mass. The largest length, which saturates this inequality, enables us to define an energy density corresponding to the holographic principle as  \cite{WANG20171},
	\begin{equation}
	\label{eqn4}
	\rho_{_\Lambda}=3c^{2}M_{p}^{2}L^{-2}
	\end{equation}
where $c^{2}$ is the model parameter. The energy density $\rho_{_\Lambda}$ can be adopted as the dark energy density in the cosmological context, with $L,$ the IR cutoff, as the cosmological length scale. There are various approaches to choose the length scale, in which the most straightforward choice is the Hubble horizon, $ L=1/H$. Even though this assumption accounts for the present dark energy density of the universe \cite{WANG20171}, it failed badly in explaining the current accelerated expansion of the universe with non-interacting dark sectors since it predicts an equation of state around $\omega_{_\Lambda}=0$ \cite{HSU200413}, which is in dispute with the condition corresponding to accelerated expansion, $\omega_{_\Lambda} < -1/3.$ However, using this IR cutoff and assuming a possible interaction between the dark sectors,  can solve the coincidence problem \cite{PAVON2005206,Sheykhi_2009,SEN20087,Pavonarticle,sadjadi2011cosmic}. Alternatively, particle horizon or event horizon becomes the reasonable choice for the cosmological length scale. On consideration of particle horizon as the length scale, the corresponding model results in an equation of state parameter $\omega_{_\Lambda} > -1/3$, thus failing to explain accelerated expansion\cite{gong2004extended}. The choice of the future event horizon as the length scale results in a model which predicts the recent acceleration of the universe and is also favored by the observational data to some extent\cite{fischler1998holography,PhysRevD.72.043510,Cruz2020ModelingHD}. However, it faces the causality problem, such that the nature of present dark energy seems to depend on the future evolution of the scale factor\cite{LI20041}.
 
Later attention has then turned to defining the holographic dark energy with curvature scalar as the IR cutoff\cite{George.2018myt,zhang2009holographic,fu2012holographic}. Following this, as a formal generalization of the Ricci scalar curvature, a hybrid cut-off scale known as the Granda-Oliveros (GO) scale has been proposed\cite{Granda_2008}. The GO cut-off scale, defined as the combination of the Hubble parameter and its time derivative, is used to formulate holographic dark energy density. The corresponding model thus depends on the local quantities; hence, no causality problems arise. Much interest has arisen in holographic dark energy with GO cut-off\cite{GRANDA2009199,karami2010new,PhysRevD.89.123009,khodam2013statefinder,pasqua2016powerlaw,malekjani2011cosmological,korunur2019tsallis,Oliveros:2014kla}. Recently a new model of holographic dark energy, known as the Barrow holographic dark energy, has been proposed based on the Barrow entropy. Barrow has expressed that the black hole surface has an intricate, fractal structure due to quantum gravitational effects, which leads to a finite volume but infinite (or finite) area. This distorted geometrical structure could have implications on the form of horizon entropy. Following this perspective, Barrow has proposed that the black hole entropy deformed to a more general relation \cite{BARROW2020135643},
\begin{equation}
\label{eqn7}
S_{h} = \left(\frac{A}{A_{0}}\right)^{1+\Delta/2}
\end{equation} 
where $A$ and $A_{0}$ are the area of the black hole horizon and Planck area, respectively. The exponent $\Delta$ quantifies the amount of quantum-gravitational deformation effects. In literature the Barrow exponent (or Barrow index), takes the range $0\leq \Delta \leq 1 $. The upper limit of $\Delta$ corresponds to the most intricate and fractal structure of the black hole horizon, while the lower limit, $\Delta =0$ corresponds to the simplest horizon structure, at which it reduces to the standard Bekenstein-Hawking entropy \cite{Bekenstein:1972tm,PhysRevD.7.2333,PhysRevD.49.1912}. The above deformation in entropy differs from the logarithmic corrections to the black hole entropy \cite{Carlip_2000,PhysRevLett.84.5255}. Even though the above expression bears some resemblance to Tsallis non-extensive entropy, the underlying principles are completely different \cite{Tsallis_2013,PhysRevLett.84.2770}.  In reference\cite{PhysRevD.102.123525}, the author proposed a specific holographic dark energy model by employing the holographic principle with Barrow entropy as the horizon entropy. The standard holographic principle can also be stated as $\rho_{_\Lambda} L^4 \leq S.$ Using Barrow entropy in Eq.~(\ref{eqn7}), the Barrow holographic dark energy (BHDE) density can be defined as \cite{PhysRevD.102.123525},
	\begin{equation}
	\label{eqn8}
	\rho_{_\Lambda}= 3CL^{\Delta-2}
	\end{equation}
Where $C$ is the parameter with dimension $[L]^{-2-\Delta}$, the factor $3$ is used for convenience. As $\Delta=0$ then $\rho_{_\Lambda}= 3 c^{2}M_{p}^{2}L^{-2},$ which corresponds to the standard holographic dark energy with $C=c^{2} M_{p}^{2} $. Using Eq. (\ref{eqn8}) it is possible to construct various BHDE models by taking different length scales as IR cut-offs in either interacting or non-interacting scenarios. In \cite{PhysRevD.102.123525}, authors analyzed the thermal history of the flat universe having BHDE with future event horizon as IR cut-off and non-relativistic matter. They have shown that the Barrow exponent $\Delta$ has a significant role in determining the evolutionary nature of dark energy. For $\Delta>0.5$ the dark energy shows a phantom behavior. By using the data from Supernovae (SN Ia) Pantheon sample and Hubble data from Cosmic Chronometer (CC), the authors in \cite{Anagnostopoulos:2020ctz} constraint the Barrow exponent in the range 0.094 - 0.095 and inferred that non-interacting BHDE model with future event horizon as IR cut-off is compatible with standard $\Lambda$CDM and in good agreement with observational data. The generalized second law (GSL) of thermodynamics of the universe with Barrow entropy as the apparent horizon entropy has been examined,\cite{Saridakis_2021} and has shown that the GSL is always valid without quantum gravitational deformation. Furthermore, their analysis showed that the generalized second law could be conditionally violated for finite non-zero values of $\Delta$. The GSL is usually valid only for cases where the Barrow exponent should be constrained to a small value such that the Barrow entropy shows only a small deviation from the standard relation. However, assuming $\Lambda$CDM-like behavior for the Hubble parameter, the Validity of GSL can be restored independently of the $\Delta$ values. There is a claim of strong Bayesian evidence for BHDE with Hubble horizon as IR cut-off compared with the standard $\Lambda$CDM in reference\cite{Dabrowski:2020atl}. In \cite{Saridakis_2020}, the author extracted the modified Friedmann equations from the first law of thermodynamics, $-dE=TdS,$ for a (3+1)-dimensional FLRW universe with Barrow entropy for the horizon and found that, on neglecting the Barrow exponent the conventional form of Friedmann equation can be recovered. In the same paper, it is argued that the dark energy equation of state parameter shows a deviation from the $\Lambda$CDM behavior at intermediate times. However, the model predicts a de Sitter epoch independent of the Barrow exponent at asymptotically large times. Similarly, Sheykhi \cite{Sheykhi:2021fwh} obtained the modified Friedmann equation with Barrow entropy, using the unified first law of thermodynamics, $dE=TdS+WdV,$ where $W=(\rho-P)/2,$ the work density. In addition,  the author explored the generalized second law and has shown that the generalized second law of thermodynamics is valid for a universe with fractal boundaries. The interacting BHDE model with $H$ as the IR cut-off has been examined in \cite{Mamon:2020spa}, where the authors pointed out the possibility of a conditional violation of GSL. Authors in \cite{Dixit:2021phd} studied the BHDE for a non-flat FLRW universe with the apparent horizon as IR cut-off and have shown that the dark energy has quintessence behavior. However, for a flat universe, the dark energy exhibits quintessence and chaplygin gas behavior \cite{Pradhan_2021}. In reference \cite{Huang:2021zgj}, the authors have carried out the dynamical analysis and the statefinder diagnostic on interacting BHDE with different IR cut-offs. Recently, the thermodynamics of the BHDE model has been discussed with a specific cut-off, called Nojiri-Odintsov (NO) cut-off, which is a particular combination of the event horizon scale \cite{Chakraborty:2021uzp}. In the same paper, the authors incorporated the bulk viscosity in the BHDE sector and studied the evolution of the cosmological parameters. Then showed that the model satisfies the GSL. Very recently, the BHDE model has been studied in the non-flat universe and explains the thermal history of the universe with the matter, and dark energy eras \cite{Adhikary:2021xym}. They also investigated the effect of Barrow exponent $\Delta$ on the equation of state (EoS) parameter. For $\Delta=0$, the EoS parameter lies entirely in the quintessence region, while for $\Delta >0.03 $, they claimed that the model crosses the phantom divide.\\

The holographic dark energy is defined by relating the  IR cut-off, corresponding to the size of the universe, to the UV cut-off, corresponding to the scale of vacuum energy. Since the UV cut-off is related to vacuum, it is more significant to reconsider holographic dark energy as decaying vacuum energy with the equation of state parameter, $\omega_{\Lambda}=-1.$ So far, all the works in the current literature on BHDE models have been done by considering the dark energy equation of state parameter as varying with the expansion of the universe. So it's worth analyzing the consequences of treating BHDE as a decaying vacuum. Motivating from this, we treat the BHDE as a dynamical vacuum in the present work. In this study, we have accounted for the interaction between the dark sectors phenomenally. Our study provides the interesting result that, in the absence of interaction, the model shows an evolution similar to that of the $\Lambda$CDM by predicting an effective cosmological constant. However, with the presence of interaction,  our model predicts a quintessence behavior of the universe that perfectly obeys the principles of thermodynamics, particularly the GSL of thermodynamics. We extracted the model parameters using the latest cosmological data and found that the universe evolves to an end de Sitter epoch. We have also performed the dynamical system analysis and found that the end de Sitter epoch is a stable equilibrium. Later we performed the Information criteria analysis for model selection.

We organize the paper as follows. The next section introduces an interacting BHDE model with GO IR cut-off, and the corresponding Hubble parameter of the model is evaluated. The third section presents the extraction of model parameters using chi-square statistics with corresponding error bars. Section four consists of the study of the evolution of the various cosmological parameters. In section the fifth, the present model is distinguished from the standard $\Lambda$CDM model by performing statefinder diagnostics. In section six, We extended our work to study the thermodynamics of the model and analyzed the status of the GSL in this model. Later in section seven, we explored the dynamical system analysis following the  Information Criteria analysis in section eight. We conclude our work in the last section.

\section{Interacting Barrow Holographic Dark Energy}
\label{ibhde}
Consider a spatially flat, homogeneous, and isotropic Friedmann-Leima$\hat{i}$tre-Robertson-Walker(FLRW) universe with background metric given by, 
\begin{equation}
dS^2=-dt^2+a(t)^2[dr^2+r^2d\Omega^2]
\end{equation}
Where $a(t)$ is the scale factor at time $t$, $d\Omega^2=d\theta^2+\sin^2\theta d\phi^2$ and $(r,\theta,\phi)$ are the co-moving coordinates. The Friedmann equation for such a universe, with $k=0$ (flat universe) having dark matter, with density $\rho_{m},$ and dark energy, with density $\rho_{_\Lambda},$ is,
\begin{equation}
\label{eqn10}
3H^{2}=\rho_{m}+\rho_{_\Lambda},
\end{equation}
where $H=\dot a/a$ is the Hubble parameter. Here we adopt the natural units, such that $\hbar=1=c,$ $8 \pi G=1$ (unless we mentioned). In the present work, we take the dark energy density $\rho_{_\Lambda}$ as Barrow holographic dark energy (BHDE) with Granda-Oliveros (GO) length scale as IR cut-off. The GO cut-off is given by \cite{Granda_2008},
\begin{equation}
\label{eqn5}
L^{-2}=(\alpha_{1} H^{2} +\beta_{1} \dot{H}).
\end{equation}
Here $\alpha_{1}$ and $\beta_{1}$ are dimensionless parameters,  and $\dot{H}$ is the time derivative of Hubble parameter. Substituting the  GO length scale in Eq.~(\ref{eqn5}) in to Eq.~(\ref{eqn8}), we  get  the BHDE density as,
\begin{eqnarray}
\label{eqn9}
\rho_{_\Lambda}=3(\alpha H^{2} +\beta \dot{H})^{\frac{2-\Delta}{2}}
\end{eqnarray}
where $\alpha$ and $\beta$ are constant parameters with dimension $[L]^\frac{-2\Delta}{2-\Delta}$. 
In  the present study we  treat the BHDE as a dynamical vacuum, so its equation of state becomes $\omega_{_\Lambda}=-1$ (i.e. $ p_{_\Lambda}= -\rho_{_\Lambda}$).  The second cosmic component is non-relativistic matter, with equation of state, $\omega_m=0,$ (i.e. $p_{m}= 0$). We account the interaction between these sectors phenomenologically, and hence the total energy-momentum conservation can be bifurcated in to\cite{Pereira_2009}, 
\begin{equation}
\label{eqn11}
\dot{\rho}_{_\Lambda}=-Q  \quad ; \quad
\dot{\rho}_{m}+3H\rho_{m}=Q,
\end{equation} 

where $Q$ is the interaction term, characterized by the amount of energy transfer between the dark sectors. A negative value of $Q$ corresponds to an energy transfer from dark matter to dark energy, while a positive value implies an energy transfer in a reverse manner. Owing to the lack of information about the microscopic origin of the interaction between the dark sectors,  \cite{Praseetha:2015bjf,pasqua2016powerlaw,Ar_valo_2012,Wang:2005jx,Bolotin:2013jpa,Feng:2007wn,Sadeghi_2010,Sadjadi:2007ts}  we choose $Q= 3bH\rho_{m}$\cite{Sadri_2020}, where $b$ is a dimensionless coupling parameter, characterizing the strength of energy flow \cite{Pereira_2009,Mamon:2020spa}. 
Using the variable, $x= ln a$ , the above conservation law in Eq.~(\ref{eqn11}) can be reformulated as,
	\begin{equation}
	\label{eqn12}
	\frac{d\Omega_{_\Lambda}}{dx} = -3b\Omega_{m} \quad ; \quad
	\frac{d\Omega_{m}}{dx} = -3(1-b)\Omega_{m},
	\end{equation}
where $\Omega_{m}=\rho_{m}/(3H_{0}^{2})$ and $\Omega_{_{\Lambda}}=\rho_{_{\Lambda}}/(3H_{0}^{2})$ with $H_0$ is the present value of the Hubble parameter. The present values of the mass parameters, $\Omega_{_{\Lambda0}}$ and $\Omega_{m_0},$ satisfying the constrained relation,  $\Omega_{_{\Lambda0}}+\Omega_{m_0}=1$.  The second of the above equations, implies the evolution of the matter density as, $\Omega_{m}=\Omega_{m_0} e^{-3(1-b)x}.$
Now Eq.~(\ref{eqn12}) together with Eq.~(\ref{eqn10}) gives a second order differential equation for the weighted Hubble parameter,
\begin{equation}
\label{14}
\frac{d^{2}\bar{H}^{2}}{dx^{2}} + 3 \frac{d\bar{H}^{2}}{dx} = - 9b\Omega_{m}
\end{equation}
where $\bar{H}=H/H_{0},$  
On solving Eq.~(\ref{14}), we get, 
\begin{equation}
\label{15}
\bar{H}^{2}=\frac{\Omega_{m}}{1-b}-\frac{\Gamma_{1}}{3}a^{-3} +\Gamma_{2}
\end{equation}
Here $\Gamma_{1}$ and $\Gamma_{2}$ are constant parameters, which are to be evaluated using suitable conditions. For this we take both the above solution and ${d\bar{H}^2/dx}$ obtained from the relation of dark energy density in Eq.~(\ref{eqn9}), and evaluated their values at  $a=a_0=1,$ where $a_0$ is the present scale factor of expansion and is standardized as one, we then get, 
\begin{equation}
\bar{H}^{2}\mid_{_{a=1}}=1 \quad \quad  \left.\frac{d\bar{H}^{2}}{dx} \right|_{_{x=0(a=1)}}= \frac{2}{\beta}\left( (H_{0}^{\Delta} \Omega_{_{\Lambda0}})^{\frac{2}{2-\Delta}}-\alpha \right)
\end{equation}
The constants are obtained as,
\begin{equation}
\label{17}
\Gamma_{1}=3 \Omega_{m_0}+\frac{2}{\beta}\left[\left( (1-\Omega_{m_0})H_{0}^{\Delta}\right)^{\frac{2}{2-\Delta}}-\alpha\right]
\end{equation}
\begin{eqnarray}
\label{eqn18}
\Gamma_{2}= 1- \frac{b\Omega_{m_0}}{1-b} + \frac{2}{3\beta}\left[\left( (1-\Omega_{m_0})H_{0}^{\Delta}\right)^{\frac{2}{2-\Delta}}- \alpha\right]
\end{eqnarray}
Now we will analyze the features of the Hubble parameter in Eq.~(\ref{15}). 
In the asymptotic limit, as $a\rightarrow 0$,  the first two terms in Eq.~(\ref{15}) dominates over the constant $\Gamma_{2}.$ As a result the Hubble parameter satisfies,  $\bar{H}^{2} \to \left(\frac{\Omega_{m_0}}{1-b} a^{3b}-\frac{\Gamma_{1}}{3}\right)a^{-3},$ which shows a decelerated expansion for the interaction parameter in the range $0\le b<1$. For the future limit as $a\rightarrow +\infty$ , the constant $\Gamma_{2}$ became the dominated term in Eq.~(\ref{15}),  hence $ \bar{H}^{2} \rightarrow \Gamma_{2},$  which implies the end de Sitter phase. So the model predicts a transition to the late accelerated epoch from a prior matter dominated decelerated phase.
It is interesting to note that, for $b=0$, i.e. when the dark sectors are not interacting with each other, then Eq.~(\ref{15}) reduces to the form,
\begin{equation}
\label{lcdm}
\bar{H}^{2}= \bar{\Omega}_{m_0}a^{-3}+\bar{\Omega}_{_{\Lambda0}}
\end{equation}
This implies that, the present model shows a $\Lambda$CDM-like behavior, with effective mass parameters\cite{1998} for non-relativistic matter,  $\bar{\Omega}_{m_{0}}$ and dark energy  $\bar{\Omega}_{_{\Lambda0}}$ as,
\begin{equation}
\label{massparameter}
\bar{\Omega}_{m_0}=\frac{2\alpha}{3\beta}\left[ 1-\frac{\left((1-\Omega_{m_0})H_{0}^{\Delta}\right)^{\frac{2}{2-\Delta}}}{\alpha}\right], \quad \quad 
\bar{\Omega}_{_{\Lambda0}}=1-\bar{\Omega}_{m_0}
\end{equation}
respectively. The term $\bar{\Omega}_{_\Lambda{0}}$, is analogues to an effective cosmological constant.  The reason for this behavior is that for $b=0$  the conservation laws in Eq.~(\ref{eqn11}) reduce to separate conservation, i.e., matter and dark energy. Hence, they are independently conserved such that the dark energy density will effectively become a constant.

\section{Contrasting the Model with Observational Data}
\begin{figure}[h]
	\centering
	\includegraphics[width=0.750\linewidth]{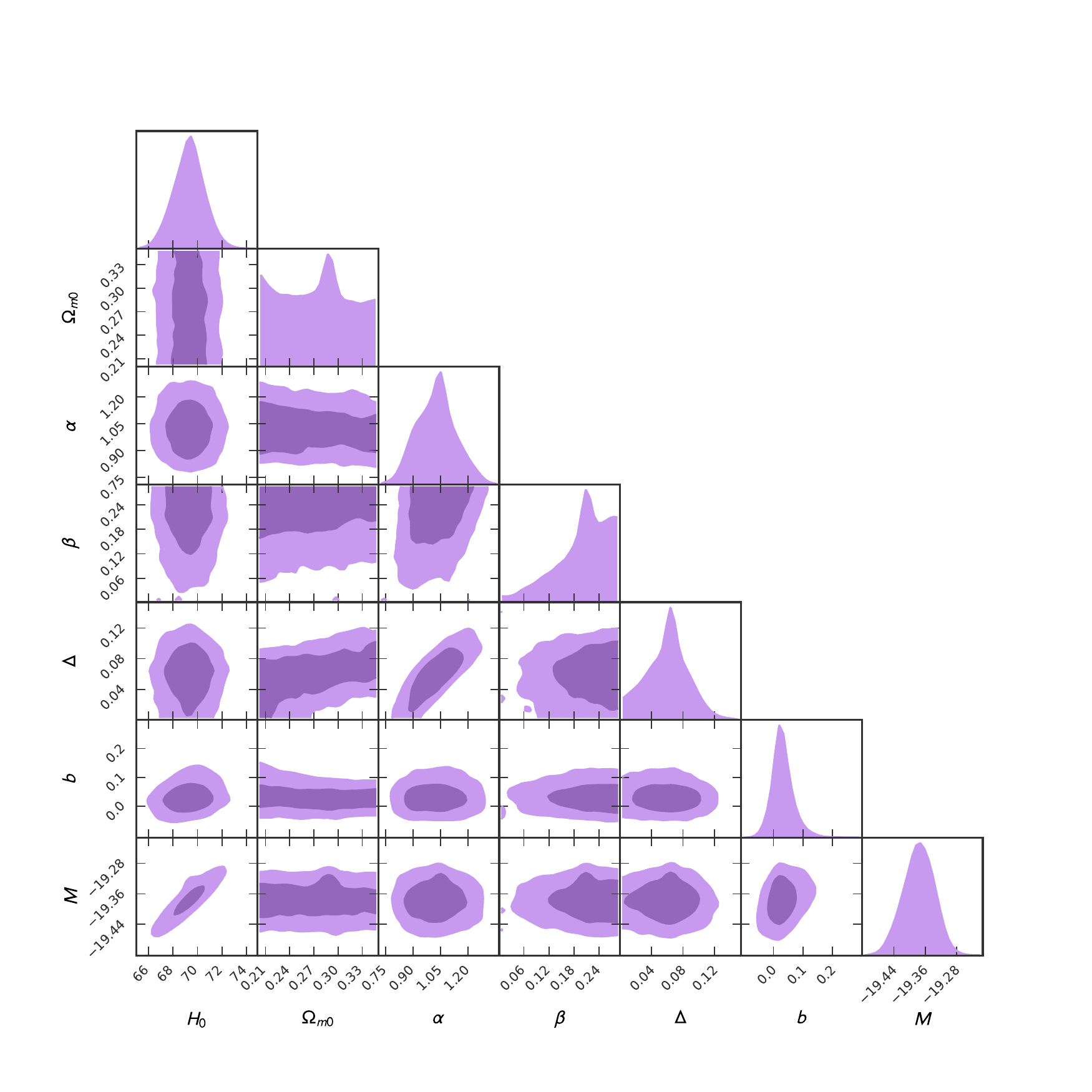}
	\caption{Corner plot of two-dimensional posterior contours with $1\sigma(68\%)$ and $2\sigma(95\%)$ confidence level and one dimensional marginalized posterior distributions of model parameters for the combined dataset $OHD36+SN Ia$}
	\label{fig:cornerplot}
\end{figure}

We adopted the chi-square ($\chi^{2}$) statistics method to obtain the best-fit model parameters. Let $\xi$ is a cosmological variable, for which, there is an observational value $\xi_{obs}$, and a corresponding theoretically predicted value,  $\xi_{theo}(P).$ The predicted value depends on the model parameters, say $P.$ In the present model the parameters are $P = H_{0},\Omega_{m_0}, \alpha, \beta, \Delta, $ and $ b $. Then the $\chi^{2} $ function defined as,
\begin{equation}
\label{eqn20}
\chi^{2}_{\xi}(P)= \sum_{i} \frac{\left[\xi_{theo}(P)-\xi_{obs}\right]^{2}}{\sigma_{\xi}^{2}}
\end{equation}
where  $\sigma_{\xi}$ is the standard deviation associated with the observation of the physical quantity. The most probable values of the model parameters can then be extracted by statistically minimizing the $\chi^2$ function. In order to perform the $\chi^{2} $ minimization, we have applied the Markov chain Monte Carlo (MCMC) algorithm by considering the emcee python package \cite{Foreman_Mackey_2013} using lmfit library \cite{newville_matthew_2014_11813}. We have used the observational Hubble dataset containing 36 H(z) measurements ($OHD36$) in the redshift range $0.07\leq z\leq 2.36$ \cite{2018ApJ...856....3Y,AMIRHASHCHI2020100557}, in which 31 data are obtained using the Cosmic Chronometric technique, 3 measurements are from the radial BAO signal in the galaxy distribution, and the last 2 are measured from the BAO signal in the Lyman forest distribution alone or cross correlated with QSOs\cite{2018ApJ...856....3Y}. For observational Hubble data, the expression for $\chi^{2}$,
\begin{equation}
\label{chiohd}
\chi^{2}_{OHD36}(H_{0},\Omega_{m_0}, \alpha, \beta, \Delta, b )= \sum_{i=1}^{36} \frac{\left[H(H_{0},\Omega_{m_0}, \alpha, \beta, \Delta, b, z_i)-H_{i}\right]^{2}}{\sigma_{i}^{2}}
\end{equation} 
where $H(H_{0},\Omega_{m_0}, \alpha, \beta, \Delta, b, z_i)$ and $H_{i}$ are the theoretical and observed Hubble parameter respectively and $\sigma_{i}$ is the corresponding standard deviation.
Another set of data is taken from the frequently used cosmological probe, type Ia Supernovae (SN Ia). We have used the SN Ia Pantheon dataset with 1048 apparent magnitude ($m$) versus redshift measurements in the redshift range $0.01 \leq z \leq 2.3$ \cite{2018ApJ...859..101S}.
 The $\chi^2$ function corresponding to the Supernovae data can be expressed as,
\begin{equation}
\label{chisn}
\chi^{2}_{SN Ia}(H_{0},\Omega_{m_0}, \alpha, \beta, \Delta, b, M )= \sum_{i=1}^{1048}\frac{\left[m(H_{0},\Omega_{m_0}, \alpha, \beta, \Delta, b, M,  z_{i})-m_{i}\right]^{2}}{\sigma_{i}^{2}}
\end{equation}
where $m(H_{0},\Omega_{m_0}, \alpha, \beta, \Delta, b, M, z_{i})$ is the theoretical value of the apparent magnitude of the Supernovae at a given redshift and $m_i$ is the corresponding observed magnitude. The $\sigma_i$ in the denominator of the above equation is now the standard deviation in the observation of the Supernovae. The theoretical magnitude is evaluated by determining the luminosity distance relation,
\begin{equation}
d_{L}(H_{0},\Omega_{m_0}, \alpha, \beta, \Delta, b, z_{i})=c(1+z_{i})\int_{0}^{z_{i}}\frac{dz}{H(H_{0},\Omega_{m_0}, \alpha, \beta, \Delta, b, z)}.
\end{equation}
 Thus, the apparent magnitude can be determined by the standard relation, 
\begin{equation}
m(H_{0},\Omega_{m_0}, \alpha, \beta, \Delta, b, M, z_{i}) = M+5log_{10}\left(\frac{d_{L}(H_{0},\Omega_{m_0}, \alpha, \beta, \Delta, b, z_{i})}{Mpc}\right)+25,
\end{equation}
 where $M$ is absolute magnitude of the supernovae is usually treated as a nuisance parameter. The most probable values of the model free parameters is then estimated corresponds to the minimal $\chi_{total}^{2}$ value of the combined dataset of observational Hubble data and Supernovae type Ia ($OHD36+SN Ia$) as, $\chi_{total}^{2}=\chi^{2}_{OHD36}+\chi^{2}_{SN Ia} $ with $\chi^{2}_{min}$ =$1053.927 $ and $\chi^{2}_{d.o.f}$ = $0.979$. The absolute magnitude $M$ is extracted as $ -19.378\pm 0.039$. By using the pygtc open python package\cite{Bocquet2016}, the plot of two-dimensional posterior contours with $1\sigma(68\%)$ and $2\sigma(95\%)$ confidence level and one dimensional marginalized posterior distributions of model parameters $ H_{0},\hspace{0.1cm}  \Omega_{m_0},\hspace{0.1cm}  \alpha,\hspace{0.1cm}  \beta,\hspace{0.1cm}  \Delta$,  $b $ and $M$ are performed for the combined dataset $OHD36+SN Ia$ is given in Fig.~\ref{fig:cornerplot}. The best-estimated values of various model parameters for 1$\sigma$ error bar is given in Table. \ref{table1}. 
  
The Present value of Hubble parameter is extracted as $69.256\pm{1.228}$ $kms^{-1}Mpc^{-1}$ which shows a slight variation from the observational value (Planck 2018 observation 67.4$\pm$ 0.5 $kms^{-1}Mpc^{-1}$\cite{2020}, and WMAP sky survey value 71.9$^{+2.6}_{-2.7}$ $kms^{-1}Mpc^{-1}$ \cite{2009}. In a previous work\cite{PhysRevD.81.083523}, the best fit value of $H_0,$ for non-interacting HDE model with GO cut-off is estimated as $66.35^{+{2.38}}_{{-2.14}}.$ The improved value of $H_0$ in the current model shows its significance, in the light of observational results. In the same reference the estimated value of the model parameters $\alpha= 0.8502^{+{0.0984}}_{{-0.0875}}$ and $\beta= 0.4817^{+{0.0842}}_{{-0.0773}}$ \cite{PhysRevD.81.083523} are comparatively less than that of our model (ref.Table.\ref{table1}). Recently,  Oliveros et al.\cite{oliveros2022barrow} studied a non-interacting BHDE with GO cut-off and reported the parameter values, $\alpha=1.00^{+{0.02}}_{-0.02}, \beta= 0.69^{+0.03}_{-0.02}$ and $\Delta=0.000^{+0.004}_{-0.000}$, for OHD dataset \cite{oliveros2022barrow}. 
\\
In contrast to this, our model  favors a non-zero value for the Barrow index, $\Delta,$ owing to the inclusion of the possible interaction between the dark sectors. The positive value of the interaction implies an energy flow from BHDE to dark matter. This flow of energy between the components of one and the same universe is in line with the LeChatelier-Braun Principle \cite{Pav_n_2008}. In this juncture, it may be interesting to note about a similar study in Tsallis HDE model with GO cut-off \cite{Dheepika2022}, in which the authors have estimated the model parameters  as, $\alpha=0.960^{+0.138}_{-0.125}, \beta=0.337^{+0.184}_{-0.207}$, and the coupling parameter $b=0.023^{+0.033}_{-0.026}$, using dataset SNIa+OHD.

	\begin{table}

	\begin{center}
		
		\caption{Observational constraints on model parameters using $OHD36+SN Ia$ dataset}
		
		\label{table1}
		
		\begin{tabular}{l c c c c r}
			
			\hline \\ 
			$H_{0}$ \scriptsize{$kms^{-1}Mpc^{-1}$} &$\Omega{m_0}$ &$\alpha$  &$\beta$  & $\Delta$  &$b$ \\ \hline \hline \\
			$69.256^{+{1.228}}_{-1.228}$ 		&$0.281^{+{0.050}}_{-0.050}$   	&$1.030^{+{0.116}}_{-0.116}$ 	& $0.211^{+{0.063}}_{-0.063}$ &$0.063^{+{0.029}}_{-0.029}$   & $0.026^{+{0.030}}_{-0.030} $ \\ \\ \hline \\
		\end{tabular} 
	\end{center}
\end{table}

Using the best-fit model parameters, we have compared the theoretically estimated apparent magnitude with the corresponding observational magnitude, as shown in Fig.~\ref{fig:apparentmagnitude}. We have also compared the predicted Hubble parameter values with the observed one in Fig.~\ref{errorbar}. Both figures show good agreement between the prediction and observation.\\	
\begin{minipage}{\linewidth}
	\centering
	\begin{minipage}{0.45\linewidth}
		\begin{figure}[H]
			\includegraphics[width=\linewidth]{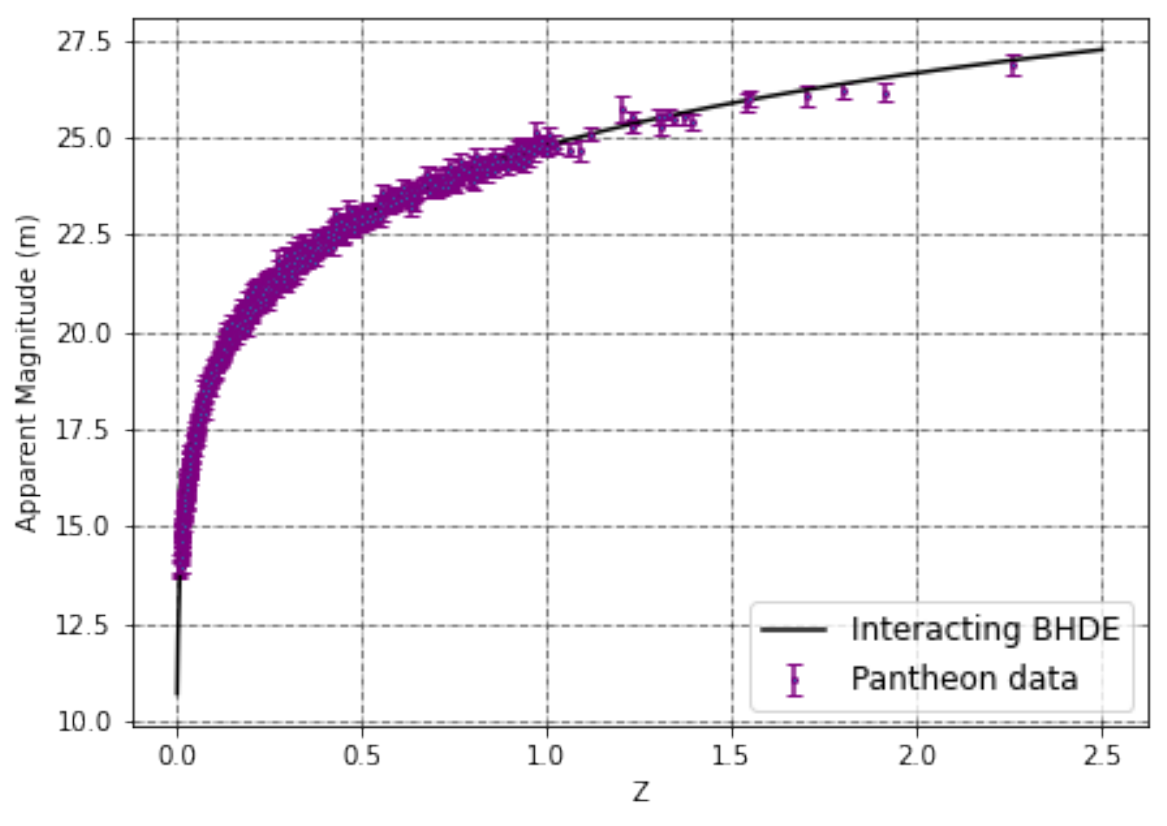}
			\caption{Comparison plot of apparent magnitude m(z) for interacting BHDE model with the best-estimated model parameters from the combined dataset $OHD36+SN Ia$ and observational Supernovae data. }
			\label{fig:apparentmagnitude}
		\end{figure}
	\end{minipage}
	\hspace{0.05\linewidth}
	\begin{minipage}{0.45\linewidth}
		\begin{figure}[H]
			\includegraphics[width=\linewidth]{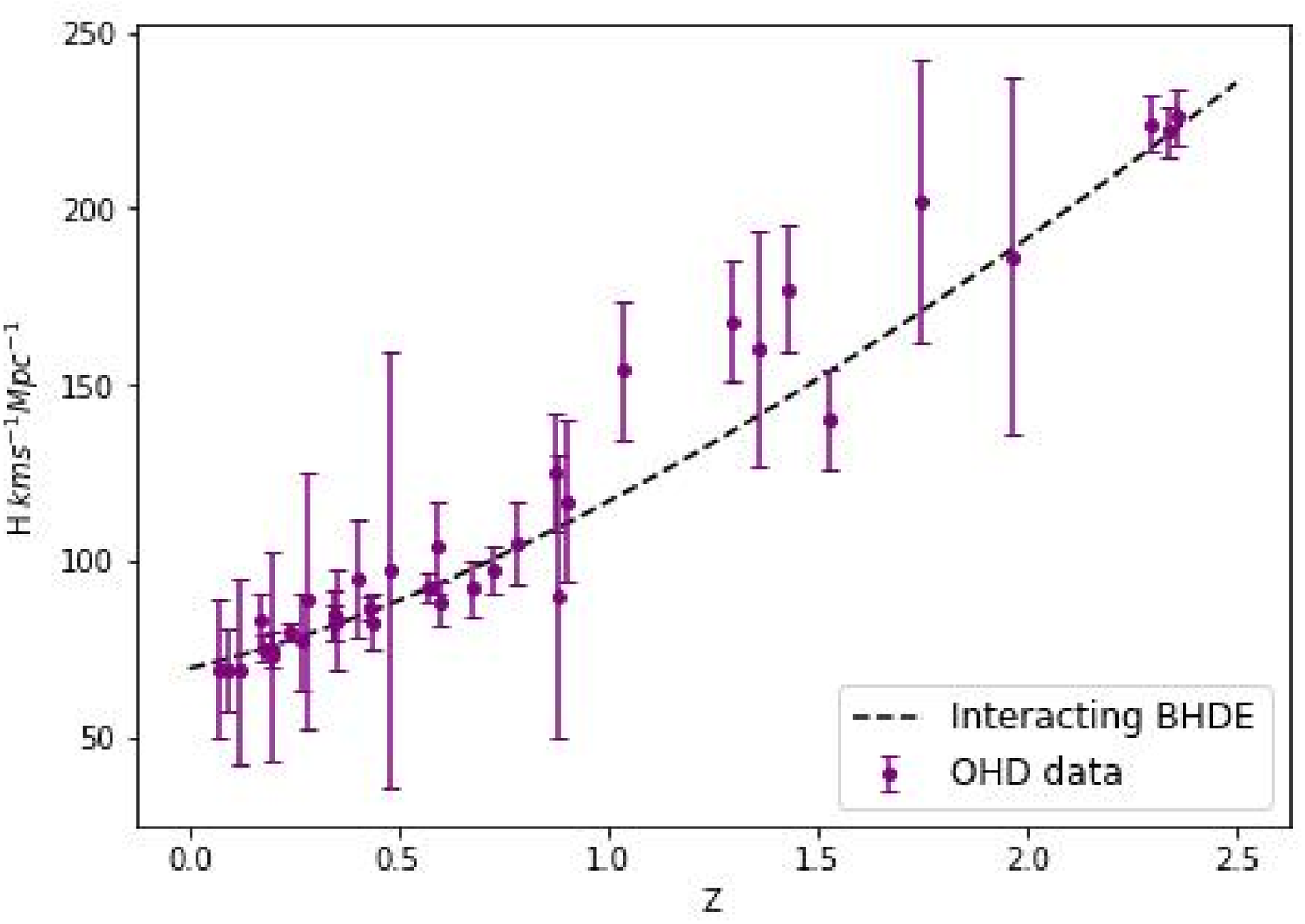}
			\caption{Comparison plot of  Hubble parameter for interacting BHDE model with best-estimated model parameters from the combined dataset $OHD36+SN Ia$ and observational Hubble data.}
			\label{errorbar}
		\end{figure}
	\end{minipage}
\end{minipage}

\section{Behavior of Cosmological Parameters}

The Eq.~(\ref{15}) shows the evolution of the Hubble parameter for the interacting BHDE model with respect to the scale factor $a$. The background evolution of Hubble parameter as a function of redshift z (where $a=1/(1+z)$) is plotted, for the best-estimated model parameters in Fig.~\ref{fig:H(z)} reveals that the evolution ultimately leads to an end de Sitter epoch.

\begin{minipage}{\linewidth}
	\centering
	\begin{minipage}{0.45\linewidth}
		\begin{figure}[H]
			\includegraphics[width=\linewidth]{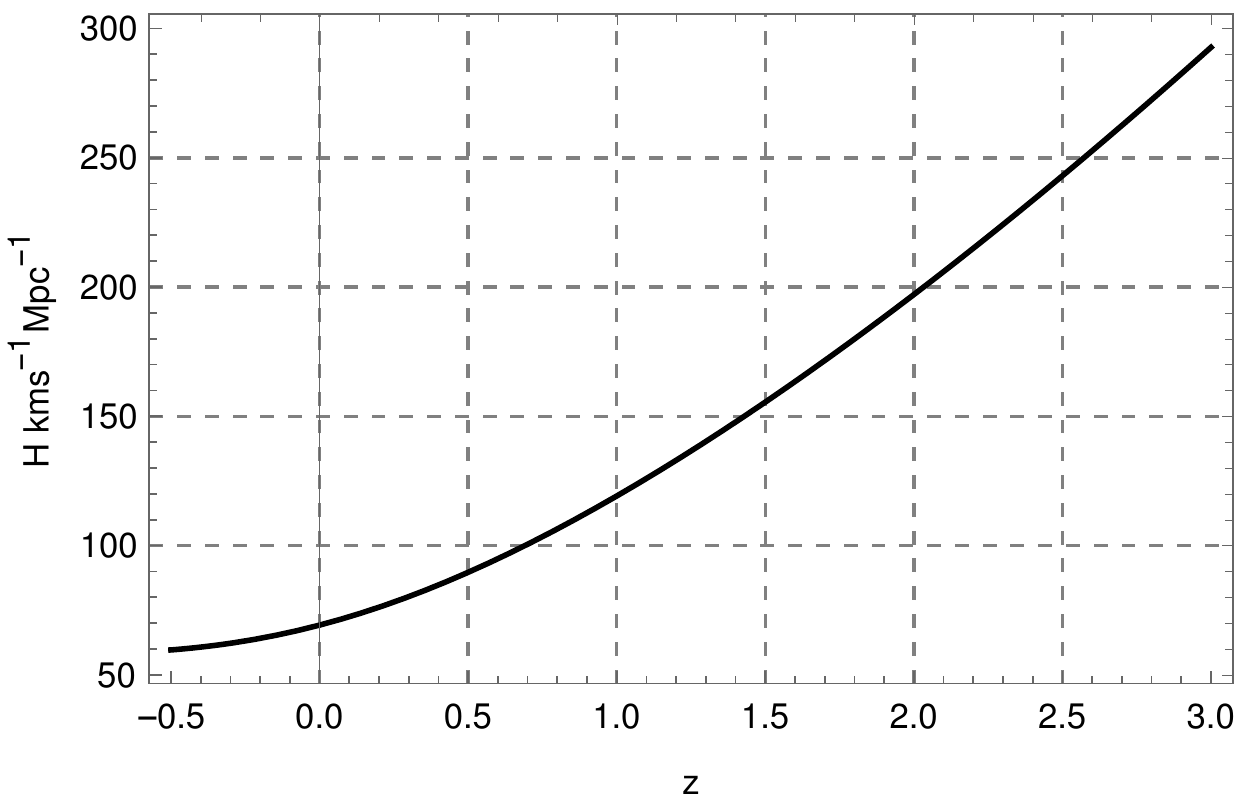}
			\caption{Evolution of Hubble parameter for the model with best-fit model parameters.}
			\label{fig:H(z)}
		\end{figure}
	\end{minipage}
	\hspace{0.05\linewidth}
	\begin{minipage}{0.45\linewidth}
		\begin{figure}[H]
			\includegraphics[width=\linewidth]{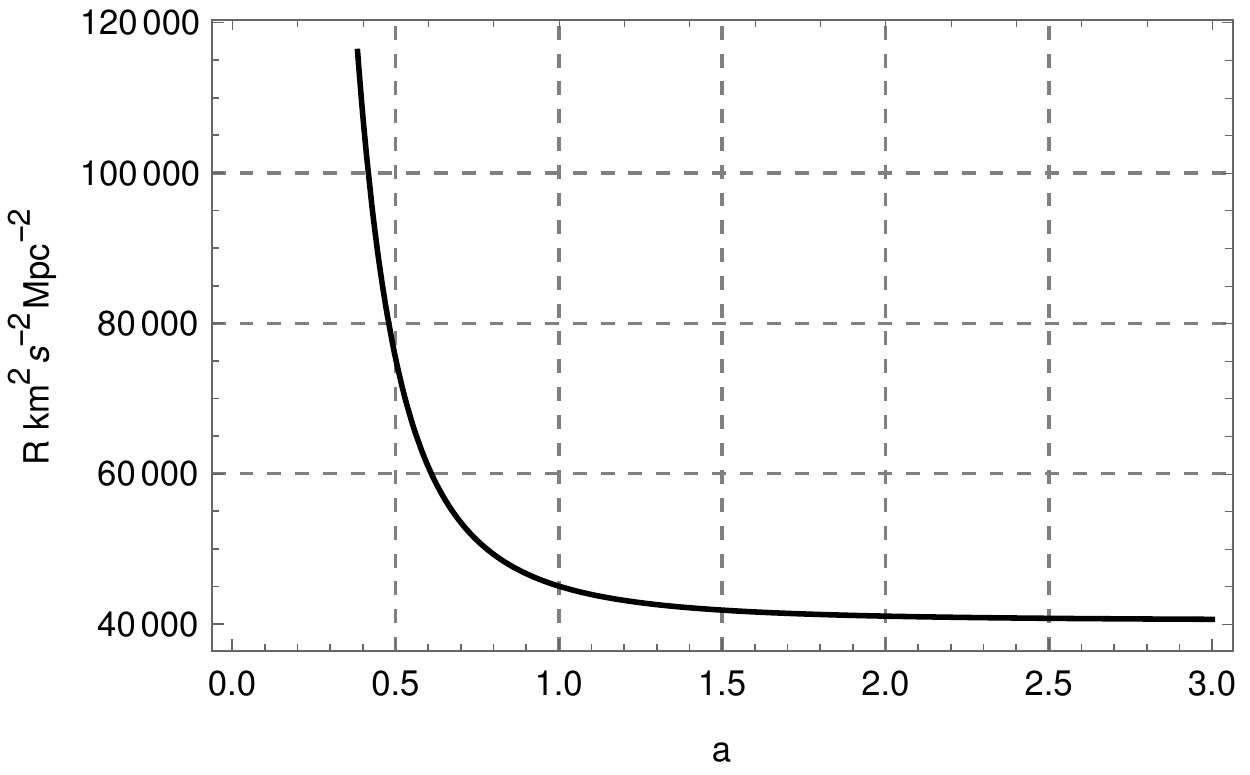}
			\caption{Evolution of curvature scalar with scale factor for the best estimate parameters}
			\label{fig:bhdecurvaturescalarpantheonohd}
		\end{figure}
	\end{minipage}
\end{minipage}
\\\\
The curvature scalar $R$ is useful for probing the early epoch of universe and is defined in terms of Hubble parameter \cite{Kolb:1990vq},
\begin{equation}
R=6(\dot{H}+2H^{2})
\end{equation}

\begin{figure}[h]
	\centering
	\includegraphics[width=0.5\linewidth]{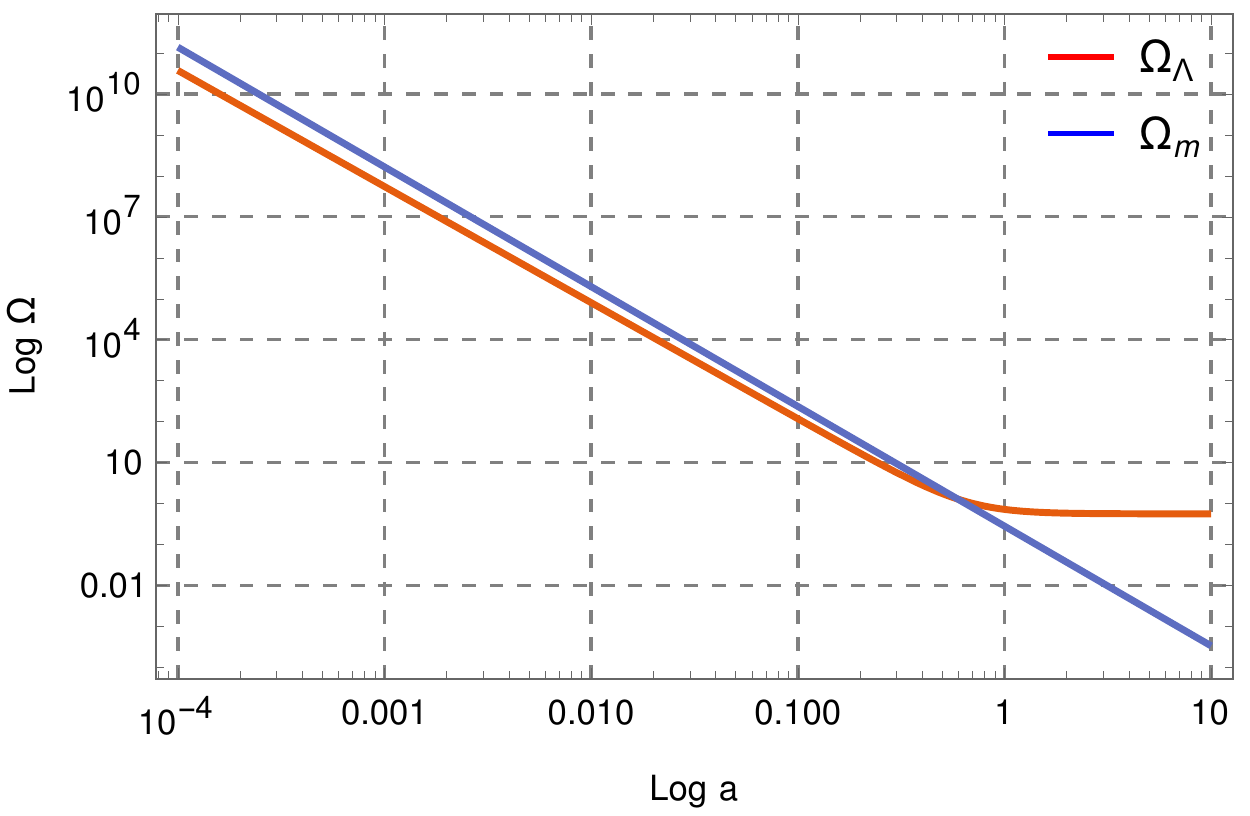}
	\caption{Evolution of density parameter corresponds to non-relativistic matter and interacting Barrow holographic dark energy in logarithmic scale.}
	\label{fig:bhdedensitylogpantheonohd}
\end{figure}

\begin{figure}
	\centering
	\includegraphics[width=0.5\linewidth]{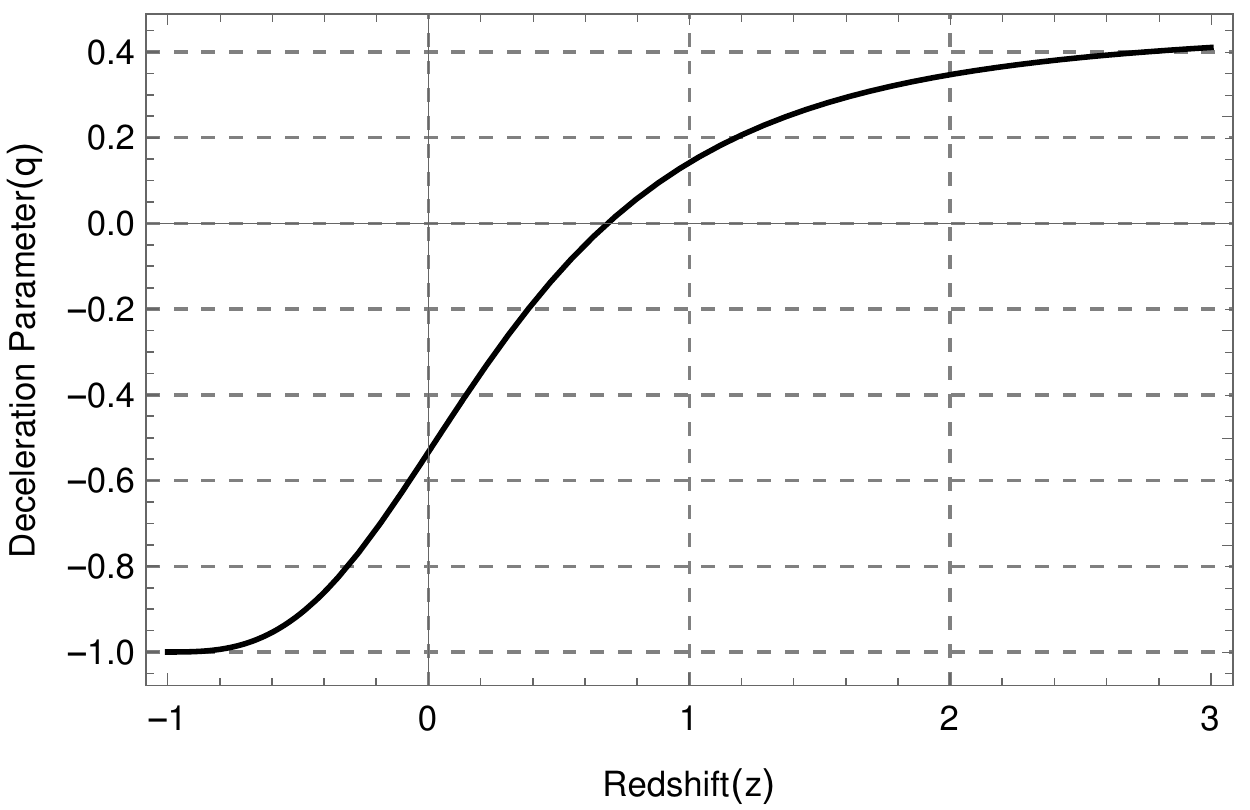}
	\caption{Behavior of q parameter vs z with transition redshift $z_{T}=0.660$}
	\label{fig:deceleration parameter}
\end{figure}

The evolution of curvature scalar parameter is as shown in Fig.~\ref{fig:bhdecurvaturescalarpantheonohd}. It is evident that, as $a\rightarrow 0$, curvature scalar approaches to infinity, which indicates the existence of the Big-bang singularity.\\
Now we will determine the age of the universe in the present model. The age obtained using the standard relation,
\begin{equation}
\label{eqn24}
t_{0}-t_{B}=\int_{0}^{1}\frac{1}{a H(a)}da
\end{equation}
Where $t_{0}$ is the present time, $t_{B}$ is the big bang time, which can be set to zero by convention. Evaluating the above integral by substituting the Hubble parameter of the present model, we found that the age of the universe corresponds to the best-estimated
 parameters is around $ 13.958$ Gyr. This is compatible with the various standard determination of age of the universe, like from WMAP data analysis, $13.70{\pm{0.2}}$ Gyr \cite{WMAP:2003elm}, $13.72 \pm0.12$ Gyr from the combined WMAP+BAO+SN Ia analysis\cite{2009} and $14\pm{0.5}$ Gyr \cite{Knox_2001} from CMB anisotropic data. Also not too much different from the age obtained from the study based on the 42 high redshift Supernovae, $14.90^{+1.4}_{-1.1}$ Gyr in reference\cite{SupernovaCosmologyProject:1998vns}.  It is to be noted that the extracted age in the present model is near to the upper age limit, 14.56 Gyr, set by observation of the oldest globular clusters. Also, our result above the lower limit proposed around 12.07 Gyr \cite{Chaboyer957}. The current model is also explain the coincidence problem. The evolution of both matter density and dark energy density are obtained as follows,  
\begin{equation}
\label{eqn25}
\Omega_{m}= \Omega_{m_0} e^{-3(1-b)x},  \quad \Omega_{_\Lambda}=\frac{1}{H_{0}^{2}}\left(\alpha H^{2}+\frac{\beta}{2}\frac{dH^{2}}{dx}\right)^{\frac{2-\Delta}{2}}
\end{equation}
The co-evolution of these mass density parameters in logarithmic scale is shown in Fig.~\ref{fig:bhdedensitylogpantheonohd}. It shows the dominance of matter over dark energy density prior to the transition and the dominance of dark energy in the later epoch. This adequately solves the coincidence problem. The present value of matter density parameter is estimated as $0.281^{+{0.050}}_{-0.050}$ which is agreeing with the current observational results, $\Omega_{m_{0}}$ $\sim 0.234 \pm 0.035- 0.29 \pm 0.007$\cite{WMAP:2006bqn}. Additionally it is in agreement with the estimated value $\Omega_{m_{0}}\sim 0.28$ for the combined dataset of Supernovae type Ia and  Cosmic Chronometer for a non-interacting BHDE model with future event horizon as IR cut-off \cite{Anagnostopoulos:2020ctz}. \\
The evolution of the deceleration parameter in the model is presented below. The deceleration parameter $q$, describes the nature of cosmic expansion, whether it is decelerating or accelerating. The evolution of this parameter can be obtained using the expression,
\begin{equation}
\label{eqn21}
q = -1-\frac{\dot{H}}{H^{2}}
\end{equation}
Precisely the expansion will be accelerating for $q<0$ and it is decelerating if $q>0$. After substituting the Hubble parameter, $q$ takes the form, 
\begin{equation}
\label{eqn22}
q=-1+\frac{ 3 \Omega_{m_0}a^{-3(1-b)}-\Gamma_{1}a^{-3}}{2\left(\frac{\Omega_{m_0}}{1-b}a^{-3(1-b)}-\frac{\Gamma_{1}}{3}a^{-3}+\Gamma_{2}\right)}.
\end{equation}
As $a \rightarrow 0$, the second term on the r.h.s. of the above equation dominates hence $q>0,$ implies the prior deceleration era, while the second term vanishes as $a \rightarrow \infty$, hence $q\rightarrow-1$  corresponding to the end de Sitter epoch. For $a= 1 $, we get the present value, $q_0$ of the parameter and it turns out that, the universe is accelerating for the current epoch, thus

\begin{equation}
q_{0}=-1+\frac{ 3 \Omega_{m_0}-\Gamma_{1}}{2\left(\frac{\Omega_{m_0}}{1-b}-\frac{\Gamma_{1}}{3}+\Gamma_{2}\right)} 
\end{equation}
in the present model. The evolutionary behavior of the deceleration parameter $q$ as a function of redshift $z$ is plotted for the best-fit model parameters obtained from $OHD36+SN Ia$ dataset, in Fig.~\ref{fig:deceleration parameter}. The value of deceleration parameter for the current epoch ($ z=0 $) is estimated as $- 0.533 $,  which is in close agreement with the observational results $q_{0}=-0.63\pm 0.12$\cite{Alam:2004jy}, $-0.644 \pm 0.22$ for latest union2SN data \cite{LI20111} and  $-0.48\pm0.11$ for SN192 sample, $-0.65\pm0.53$  for  the radio
galaxy sample \cite{daly2009cosmological}.  The transition redshift characterizing, $q=0,$ the onset of the late acceleration is found to be around $z_{T}=0.660$ and is comparable with the observational constraint using SN Ia+CMB+LSS joint analysis $z_{T}= 0.61$\cite{2018mamon} and 0.72 $\pm$ 0.05 \cite{2017} from  SN Ia+BAO/CMB dataset. But slightly higher compared to constraints using SN Ia+CMB data, $z_{T}$=0.57$\pm$0.07 \cite{Alam:2004jy}. However it should be noted that, the extracted values of $q_0$ and $z_T$ in the present model are comparable to the corresponding values obtained for the standard $\Lambda$CDM model ($z_{T}$=0.74 and $ q_{0}$=-0.59) \cite{2017}.
\section{Statefinder Diagnostics}\label{rs}
As there are a variety models of dark energy existing in the current literature, it is needed to compare the model with the standard model. Sahni et al. \cite{Sahni_2003}, have introduced the statefinder diagnostics, an effective tool in contrasting a model with the standard model and also with different dark energy models in a model independent way. The statefinder parameters $ {r,s}$ are formulated from the scale factor $a(t)$ and its derivatives with respect to cosmic time up to the third order. Following the evolutionary trajectory of the model in $r-s$ plane, it can be  classified as  quintessence, if $r < 1, s > 0$  and chaplygin gas model, if $r > 1 $ and $ s < 0$ \cite{Wu:2007st,Alam_2003,Sahni_2003}. For the standard $\Lambda$CDM model, the statefinder parameters are \{ ${r,s}$\}=$\{{1,0}\}.$  The statefinder parameter pair $\{{r,s}\}$ takes the form\cite{Sahni_2003,Alam_2003}, 
 
\begin{equation}
r=\frac{\dddot{a}}{aH^{3}}
\end{equation}
\begin{equation}
s=\frac{r-1}{3(q-\frac{1}{2})}
\end{equation}
In terms of $\bar{H}$  the above equations can be expressed as,
\begin{equation}
r=1+\frac{1}{2\bar{H}^{2}}\frac{d^{2}\bar{H}^{2}}{dx^{2}}+\frac{3}{2\bar{H}^{2}}\frac{d\bar{H}^{2}}{dx}
\end{equation}
\begin{equation}
s=\frac{r-1}{{\frac{3}{2\bar{H}^{2}}}\frac{d\bar{H}^{2}}{dx}+\frac{9}{2}}
\end{equation}
\begin{figure}[h]
	\centering
	\includegraphics[width=0.5\linewidth]{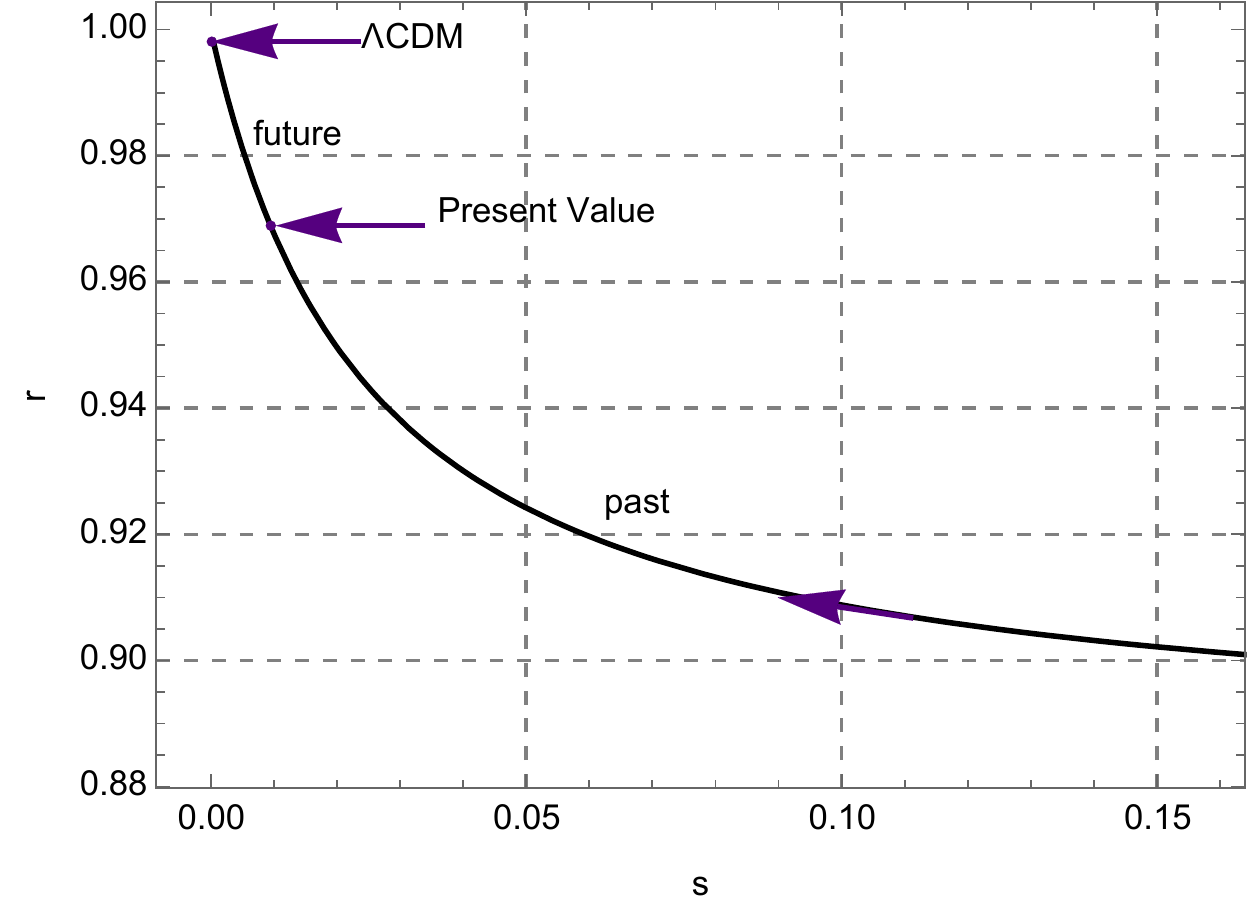}
	\caption{Evolutionary trajectory of the interacting BHDE model in r-s plane with best-estimated model parameters. Current value $\{{r_{0},s_{0}}\}$ =$\{0.968,0.010\}$}.
	\label{fig:bhdestatefinderpantheonohd}
\end{figure}
After substituting the respective terms from Eq.~(\ref{14}) and Eq.~(\ref{15}), the parameters becomes,
\begin{equation}
r=1-\frac{9b\Omega_{m_{0}}a^{-3(1-b)}+2\Gamma_{1}a^{-3})}{2\left(\frac{\Omega_{m_0}}{1-b} a^{-3(1-b)}-\frac{\Gamma_{1}}{3}a^{-3} +\Gamma_{2}\right)}
\end{equation}
\begin{equation}
s=\frac{3b\Omega_{m_{0}}a^{-3(1-b)}}{\frac{b}{1-b}\Omega_{m_{0}}a^{-3(1-b)} +\Gamma_{2}}
\end{equation}
At the asymptotic limit $a\rightarrow\infty$ the statefinder parameters takes the values, $ r \rightarrow 1$ and $s \rightarrow 0,$ shows that the present model will tends to the $\Lambda$CDM in the far future of the evolution.  The trajectory of the model in the $r-s$ plane with the best-fit parameters is depicted in Fig.~\ref{fig:bhdestatefinderpantheonohd}.
The current value of $r$ and $s$ are determined from the two-dimensional parametric plane as $\{{r_{0},s_{0}}\}$=$\{{0.968,0.010}\}$ . This trajectory in the $r-s$ plane reveals that the model is arguably different from the standard $\Lambda$CDM. It also shows that the parameters obey the condition $r < 1, s > 0$ throughout the evolution, and hence the proposed model has a quintessence behavior. \\	

\section{Thermodynamics of Interacting BHDE Model}

In this section we analyze the thermal evolution of the model, concentrating on the evolution of the entropy. The universe is evolving as an ordinary macroscopic system, hence advances towards an equilibrium state of maximum entropy. The aforementioned statement is valid only if it satisfies the appropriate constraints given by \cite{Pav_n_2012,Krishna:2017vmw,b2021emergence},
\begin{align*}
\dot{S} \geq 0 \textrm{\hspace{0.5cm}for\hspace{0.1cm} always\hspace{0.5cm}} ;\hspace{0.5cm} \ddot{S}<0 \hspace{0.5cm} \textrm{for\hspace{0.1cm} atleast \hspace{0.1cm}later\hspace{0.1cm} time\hspace{0.1cm} of\hspace{0.1cm} evolution.}
\end{align*}
where the over-dot represents the derivative with respect to cosmic time. The first one represents the condition for the validity of the generalized second law of thermodynamics, and the second one, which state that the second derivative of the entropy should be negative at least at the end stage of the universe,  is the condition to be satisfied for having an upper bound to the entropy of the system. According to GSL, total entropy comprises of matter entropy enclosed by the horizon, $S_m$, and that of horizon $S_{h},$ should always a non-decreasing function of cosmic time \cite{PhysRevD.25.942,Bekenstein:1974ax,cmp/1103858973,Bekenstein:1972tm,Pav_n_2012,PhysRevD.15.2738,Karami_2010}.
That is, 
\begin{equation}
\dot{S}_{m}+\dot{S}_{h}\geq 0
\end{equation} 
Using conservation law and integrability condition\cite{Kolb:1990vq}, the entropy becomes,
\begin{equation}
S=\frac{(\rho+P)V}{T},
\end{equation}
where $V$ is the volume of the horizon and $T$ is the temperature. The entropy of the non-relativistic matter within the horizon of volume $V=4\pi c^3 /3H^3$ is, 
\begin{equation}
S_{m}=\frac{\pi \Omega_{m}H_{0}^{2} c^{5}}{\hbar GH^{4}}k_{_B},
\end{equation}
where we took $T,$ as the Hawking temperature, $T=\hbar H/(2 \pi k_{_B}) $. For horizon entropy we have used the Barrow entropy relation such that,  
\begin{equation}
S_{h}=\left(\frac{\pi c^{5}}{\hbar G H^{2}}\right)^{1+\Delta/2} k_{B}
\end{equation}
Here $c$ is the speed of light in vacuum, $\hbar=h/2\pi$ with the Planck's constant, and $G$ is the gravitational constant. The rate of change of the total entropy with respect to the variable, $x= lna$ is given as, 
\begin{equation}
S_{tot}'=S_{m}'+S_{h}'
\end{equation}
\begin{equation}
\label{ds}
S_{tot}'= \frac{\pi H_{0}^{2} c^{5}\Omega_{m}}{\hbar G}k_{_B} \left(\frac{-3(1-b)}{H^{4}} -\frac{2}{H^{6}}\frac{dH^{2}}{dx}\right)-\left(\frac{\pi c^{5}}{\hbar G}\right)^{1+\Delta/2} \frac{2+\Delta}{2H^{4+\Delta}} \frac{dH^{2}}{dx}
\end{equation}
The evolution of $S_{tot}^{\prime}$ with scale factor is plotted in Fig.~\ref{ds1fig}, and is always positive. This shows that the entropy of the horizon plus that of matter within the horizon is non-decreasing throughout the evolution of the universe. So, the generalized second law of thermodynamics (GSL) is valid in the universe, which comprises Barrow holographic dark energy and pressure-less matter. It is to be noted that some recent works claim a conditional violation of the GSL in the BHDE model for non-zero Barrow exponent, $\Delta.$   In reference \cite{Mamon:2020spa}, the authors have considered a spatially flat FLRW universe, which constitutes pressure-less matter and BHDE that interact with each other. The authors extracted the time variation of total entropy and figure out the possibility of conditional violation of GSL. Such a conditional violation of the generalized second law is noted in reference \cite{Saridakis_2021}, in which the authors investigated the validity of GSL by considering Barrow entropy for the apparent horizon along with the entropy contributions from matter and dark energy fluids. They found that GSL may be conditionally violated depending on the evolution of the universe. If the background evolution of the Hubble function is similar to that in $\Lambda$CDM cosmology, then the generalized second law is always satisfied irrespective of the value of $\Delta$. Otherwise, the GSL be satisfied only for relatively low values of the Barrow exponent, $\Delta$ or it violates the laws of thermodynamics for a large value of $\Delta.$ So, the Barrow exponent is constrained such that the corresponding entropy is very close to the standard Bekenstein entropy. For $b=0,$ that is, dark sectors are non-interacting with each other, the Hubble parameter in the present model behave like that of the $\Lambda$CDM with the extracted mass density parameters, $\bar{\Omega}_{m_0}=0.290$ and $\bar{\Omega}_{_{\Lambda0}}=0.710$. In this situation, the GSL is perfectly valid irrespective of the value of $\Delta.$  For $b \neq 0,$ we estimated model parameters and found that the present model satisfies the GSL with the estimated model parameters. However, it should be noted that the extracted value of the Barrow exponent,  $\Delta=0.063\pm 0.029,$ is much less. Hence, the conclusion drawn in reference\cite{Saridakis_2021} that the GSL is generally valid for only low values of the Barrow exponent is true. In this line, It has been reported that the generalized second law is valid for interacting BHDE with Hubble horizon and event horizon as IR cut-off in DGP (Dvali-Gabadadze-porrati) braneworld cosmology \cite{aarticle}. Also, GSL was found to be valid in Barrow holographic dark energy with NO (Nojiri-Odintsov) cut-off \cite{sym13040562}.

 We further extended our study to check whether the present model predicts a universe that evolves to a maximum entropy state corresponding to an upper bound to the entropy at the end stage of the evolution. To check this, we have obtained the second derivative of the total entropy as,
\begin{multline}
\label{s''}
S_{tot}''=\frac{3\pi H_{0}^{2} c^{5}\Omega_{m}}{\hbar GH^{4}}k_{_B}\left[3(1-b)^{2}+\frac{4(1-b)}{H^2}\frac{dH^{2}}{dx}-\frac{2}{3H^2}\frac{d^{2}H^{2}}{dx^{2}}+\frac{2}{H^{4}}\left(\frac{dH^{2}}{dx}\right)^2\right]\\+ \left(\frac{\pi c^{5}}{\hbar G}\right)^{1+\Delta/2}\frac{2+\Delta}{2}\left[\frac{4+\Delta} {2H^{(6+\Delta)}}\left(\frac{dH^{2}}{dx}\right)^2-\frac{1}{H^{(4+\Delta)}}\frac{d^{2}H^{2}}{dx^{2}}\right]
\end{multline}
Since the relation (\ref{s''}) seems to be slightly tricky to analyze, the second derivative of the total entropy for best-estimate model parameters is depicted in Fig.~\ref{ds2fig}. Following this, we obtained that $s_{tot}''<0$ at least at the later time of evolution, guarantees the convexity of the function, i.e., in an asymptotic limit $a\rightarrow \infty$ the second derivative of the entropy approaches zero from below. So the figure points out that the model satisfies the entropy maximization condition; thus, the entropy will never grow unbounded. To a great extent, the analysis shows that the universe constitutes interacting BHDE, and pressure-less matter shows a pretty good behavior with the GSL of thermodynamics as well as evolves to a maximum entropy state analogous to the ordinary macroscopic system. 

\begin{minipage}{\linewidth}
	\centering
	\begin{minipage}{0.45\linewidth}
		\begin{figure}[H]
			\includegraphics[width=\linewidth]{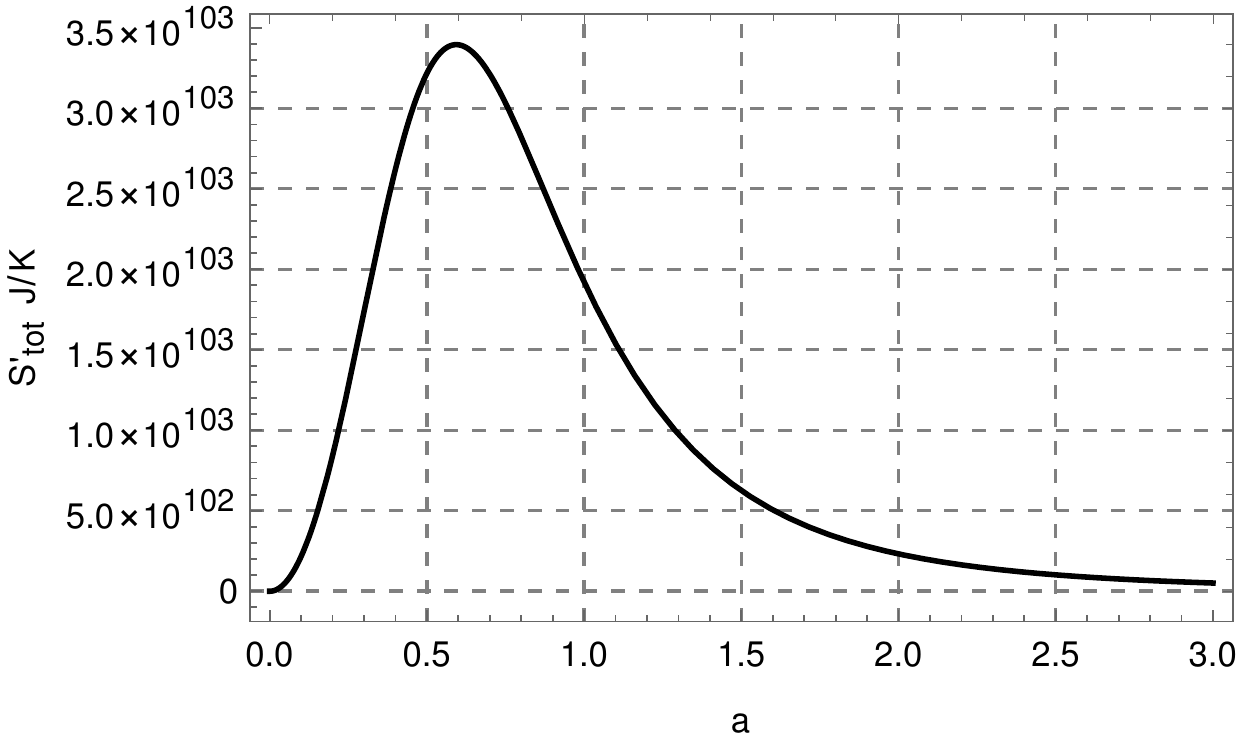}
			\caption{Evolution of $S_{tot}^{\prime}$ vs scale factor $a$. The values of model parameters is considered from the combined dataset of $OHD36+SN Ia$.}
			\label{ds1fig}
		\end{figure}
	\end{minipage}
	\hspace{0.05\linewidth}
	\begin{minipage}{0.45\linewidth}
		\begin{figure}[H]
			\includegraphics[width=\linewidth]{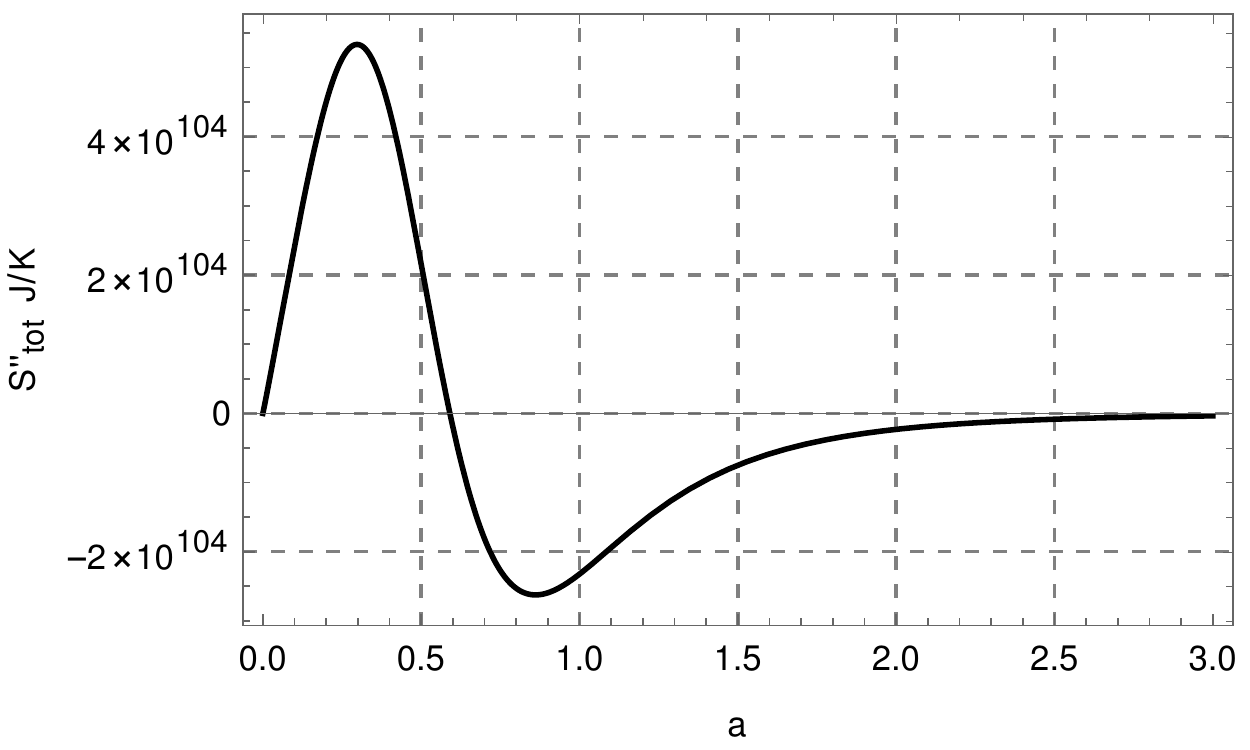}
			\caption{$S_{tot}'' $ vs scale factor $a$. The values of model parameters is considered from the combined dataset of $OHD36+SN Ia$.}
			\label{ds2fig}
		\end{figure}
	\end{minipage}
\end{minipage}
\section{Dynamical System Analysis}
\label{ps}
Phase space analysis has been widely used to obtain the asymptotic evolution of the universe in various cosmological models \cite{Coley:2003mj,mazumder2011interacting,biswas2011interacting}. The method involves the formulation of a set of autonomous differential equations of appropriately chosen dimensionless phase space variables $y_{i}$ as, 
\begin{equation}
y_{i}'=f_{i}(y_{j}) \hspace{1cm}i,j= 1,2,3,..
\end{equation}
Where the prime represents the derivative with respect to a suitably chosen variable. The equilibrium solutions are referred as the critical points ($ y_{i}=y_{i}^{*} $), obtained by considering $ y_{i}'=0 $ for all $ i $. The stability of the model corresponding to these solutions can then be obtained by analyzing the sign of the eigenvalues of the Jacobian matrix at the critical points (or equilibrium points). If the eigenvalues of the coefficient matrix is negative, the corresponding critical point is an attractor, and the neighboring trajectories will always converge to that point, thus a stable equilibrium. If eigenvalues are positive, then it is a source point where the neighboring trajectories seem to be diverging from the critical point; it is unstable. Suppose the eigenvalues have different signs, then the critical point is said to be a saddle, and the stability depends on the initial conditions\cite{gregory_2006}.

\begin{figure}
	\centering
	\includegraphics[width=0.5\linewidth]{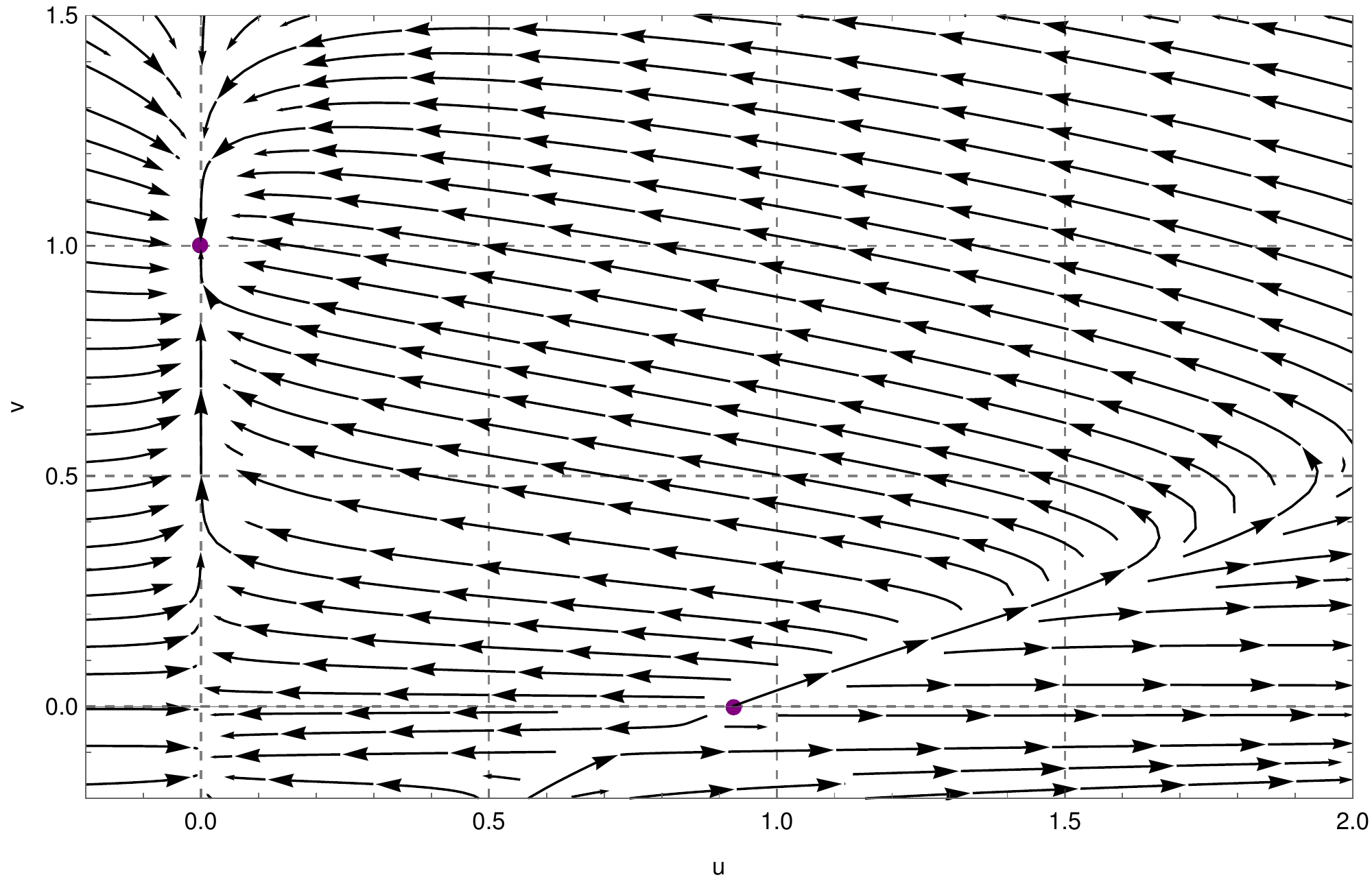}
	\caption{Phase space trajectories of the model in u-v plane with best-estimated model parameters from combined dataset $OHD36+SN Ia$.}
	\label{fig:bhdephasespace}
\end{figure}

We chose the dimensionless phase space variable $u$ and $v$ defined as,
\begin{equation}
u=\frac{\rho_{m}}{3 H^{2}} \hspace{3cm}v=\frac{\rho_{_\Lambda}}{3 H^{2}}
\end{equation}
The autonomous differential equations corresponding to these phase space variables, obtained using the Friedmann equation and the conservation principle are,
\begin{equation}
\label{eqn30}
u'=f(u,v) = 3(b - 1)u + u^{2} - 2uv + 2 u
\end{equation}
\begin{equation}
\label{eqn31}
v'= g(u,v)=\frac{(2-\Delta)v}{2}\frac{\alpha(2v-u-2)-\beta(\frac{9}{2}bu+\frac{5}{2}u+2v-2)}{ \alpha-\beta(\frac{u}{2} - v + 1)}-(2v^{2}-vu-2v)  
\end{equation}
Here the differentiation is with respect to the  variable $x=\ln a.$ By equating $u'=0$ and $v'=0,$ we obtained physically feasible critical points and are,
\begin{equation}
(u^{*},v^{*})=(0,1),(1-3b,0).
\end{equation}
In order to study the nature of these critical points, consider a small perturbation to the phase variables,  
say $u=u^{*}+\delta u$ and $v=v^{*}+\delta v,$ 
$$
\left[ \begin{array}{c} \delta u' \\ \delta v' \end{array} \right] = \begin{bmatrix} (\frac{\delta f}{\delta u})^{*} & (\frac{\delta f}{\delta v})^{*} \\ (\frac{\delta g}{\delta u})^{*} & (\frac{\delta g}{\delta v})^{*} \end{bmatrix} \times \left[ \begin{array}{c} \delta u \\ \delta v \end{array} \right]	
$$	
The $2\times2$ matrix in the above equation is the general form of the Jacobian matrix. Diagonalizing the above matrix gives the corresponding eigenvalues.  
Eigenvalues, along with the inference regarding stability conditions of these critical points, are given in Table.\ref{table2}. The phase space plot is given in Fig.~\ref{fig:bhdephasespace}.   
\begin{table}[h]
	\begin{center}

	\caption{ Critical points and Eigenvalues }
	{\begin{tabular}{@{}ccc@{}} \hline 
			Critical points \hspace{2cm} &Eigenvalues \hspace{2cm}              & Nature \\ \hline \hline \\
			$(0.923, 0)$ \hspace{2cm}		& $(0.923, 0.072)$ \hspace{2cm}		&unstable \\ \hline \\ 
			$(0, 1)$ 	\hspace{2cm}	&$(-2.926, -0.463) $ \hspace{2cm} 	&stable 	 \\ \hline \\
			
		\end{tabular} \label{table2}}
		\end{center}
\end{table}

The eigenvalues corresponding to the critical point $(0.923, 0)$ are both positive hence it is unstable. The figure reveals that trajectories around this critical point are diverging. This point, in fact, corresponds to the prior matter-dominated epoch, which is decelerating. On the other hand, the eigenvalues corresponding to the critical point $(0, 1)$ are such that both values are negative. This critical point corresponds to the dominance of Barrow holographic dark energy, hence corresponding to the end de Sitter epoch. The phase plot shows that the trajectories are converging to this critical point. Hence it can be an attractor. So the dynamical system analysis of the interacting BHDE model holds the result that the model approaches a stable end de Sitter state in the far future. A similar dynamical system analysis was performed for an interacting BHDE model with dark energy density having varying equation of state and Hubble horizon as IR cut-off \cite{Huang:2021zgj}. The authors performed the dynamical system analysis for the model with various interacting terms and figured out that the end de sitter phase is stable.

\section{Information Criteria and Model Selection}
The introduction of more model parameters improves the fitting with observational data, regardless of the relevance of extra parameters. In order to analyze the statistical significance of any model, one can perform the Information Criteria (IC) analysis such as Akaike information criterion(AIC)\cite{1100705} and Bayesian Information Criterion(BIC) or Schwarz’s Bayesian criterion  \cite{schwarz1978estimating}.  Such analysis will enable us to decide which one among the competing models is preferred by the observational data\cite{liddle2006cosmological,10.1111/j.1365-2966.2004.08033.x,10.1111/j.1745-3933.2007.00306.x}. AIC is derived by considering an approximate minimization of the Kullback-Leibler information entropy, which measures the separation between two probability distributions. The general expression for AIC is given as\cite{doi:10.1080/03610927808827599,anderson2004model,doi:10.1177/0049124104268644}, 
\begin{equation}
\label{aic}
AIC= -2 ln(\mathcal{L}_{max})+2k+\frac{2k(k+1)}{N-k-1}
\end{equation}
Where  $\mathcal{L}_{max}$ is the maximum likelihood of the considered dataset contains $N$ number of data points and $k$ is the number of model parameters. For large number of data points, the Eq.~(\ref{aic}) reduces to,
\begin{equation}
AIC= -2 ln(\mathcal{L}_{max})+2k
\end{equation}
BIC is derived by approximating the Bayes factor \cite{Jeffreys61}, defined as,
\begin{equation}
BIC= -2 ln(\mathcal{L}_{max})+kln N
\end{equation}
On interpreting the obtained values of IC components, the most admitted model with observational support is the one that minimizes the IC. Moreover, the difference in IC has more significance, is defined as,
\begin{equation}
\Delta IC_{model}=  IC_{model}- IC_{min}
\end{equation}
where $IC_{min}$ is the minimum IC value in competing models. The criteria for determining the strength of evidence against the model with minimum IC is discussed in \cite{anderson2004model,1100705,sola2017first}.  Accordingly, if $\Delta IC \le 2 $ the given model has very strong evidence against the model with minimum $IC$,   
and for the range $2-6$, there is evidence for the given model, but not strong. Suppose the parameter $\Delta IC_{model}$ is in the range $6-10$; in that case, the given model can claim only less evidence. For $\Delta$IC $>$10, the model has essentially no evidence against the model with $IC_{min}$.
We have compared the AIC and BIC values of the present interacting BHDE model with the standard $\Lambda$CDM model by evaluating the difference $\Delta AIC$ and $\Delta BIC$, and are depicted in Table.\ref{table3}. The comparison of $\Delta$IC  for both AIC and BIC analysis leads to the following conclusions. The interacting BHDE has considerably less evidence against $\Lambda$CDM in AIC analysis; however, the model has essentially no support in BIC analysis. This is because of the fact that the BIC substantially penalizes the cosmological model with a larger number of free parameters.\cite{10.1111/j.1365-2966.2004.08033.x}.
\begin{table}[h]
	\begin{center}
\caption{ Information criteria AIC and BIC of interacting BHDE model and $\Lambda$CDM model  along with the difference $\Delta$IC  }
 \label{table3}
	{\begin{tabular}{@{}ccccc@{}} \hline \\ 
					Model\hspace{2cm} & AIC  &$\Delta AIC$ &BIC  &$\Delta BIC$\\ \hline \hline \\
				Interacting BHDE\hspace{2cm}		& 1067.927		&7.457 & 1102.846 &27.411 \\ \hline \\
					$\Lambda$CDM 	\hspace{2cm}		&1060.470  	&0 	&1075.435 &0 	 \\ \hline \\
			 \end{tabular}}
\end{center} 
\end{table}

%
%

\newpage
\section{Conclusion}
The Quantum gravitational effect leads to a new area-entropy relation that deviates from the ordinary Bekenstein-Hawking relation. In this line, Barrow proposed a new relation for the horizon entropy, $S=(A/A_0)^{1+\Delta/2}.$ Following the holographic principle in cosmology, a new dark energy model, Barrow holographic dark energy, has been proposed in the recent literature. We have analyzed a cosmological model with BHDE, with the Granda-Oliveros scale as the IR cutoff,  and non-relativistic matter as components in the present work to explain the recent acceleration of the universe. The novelty is that in the present model BHDE is treated as a dynamical vacuum with a constant equation of state, $\omega_{\Lambda}=-1.$ We accounted for the interaction between BHDE and the matter by assuming a phenomenological form for the interaction. By assuming the interaction term as $Q=3bH\rho_{m},$ we have analytically solved the Friedmann equations for finding the Hubble parameter evolution. The Hubble parameter exhibits acceptable asymptotic behaviors in its evolution. As the scale factor, $a \to 0$ the Hubble parameter reduces to a suitable form, representing the prior matter dominated decelerated epoch. As $a \to \infty,$ the Hubble function tends to a constant, implies an end de Sitter epoch.   Interestingly, the present model shows an exact $\Lambda$CDM like behavior for the interaction parameter $b=0,$ in which the model predicts effective mass density parameters for both matter and dark energy. 

We extracted best-fit values of the model parameters at the $1 \sigma$ confidence interval by contrasting the model with the cosmological observational dataset $OHD36+SN Ia$.
The evolution of cosmological parameters such as Hubble parameter, curvature scalar, density, deceleration parameter, etc., have been analyzed. The obtained present value of Hubble parameter is 	$69.256^{+{1.228}}_{-1.228}$  $kms^{-1}Mpc^{-1}$, present value of matter density parameter $\Omega_{m_{0}}=0.281^{+{0.050}}_{-0.050}$ and the age of the universe is around $ 13.958$ Gyr. These are in concordance with the observational results. The behavior of the density parameter solves the coincidence problem. The transition from a prior decelerated phase to an accelerating phase is found to occur at transition redshift $z_{T}= 0.660$, and the present value of deceleration parameter $q_{0}=- 0.533 $ and is moderately consistent with the observational results.

The thermal evolution of the model has been analyzed. We found that the model satisfies the generalized second law of thermodynamics and the condition for entropy maximization. Hence the end de Sitter turns out to be a state of maximum entropy. Regarding the thermal behavior, it is to be noted that if there is no interaction between the dark sectors, the model predicts an evolution exactly similar to the standard $\Lambda$CDM and hence the validity of GSL is automatically satisfied, for any value of the Barrow index, $\Delta.$ It has been found that, with the interaction between the dark sectors also, the GSL is satisfied. However, this time, it seems that the validity of the GSL is guaranteed due to the low value of the $\Delta$ parameter extracted in comparison of the model with the observational data. In this regard, the results of our model is in line with an earlier results in the literature that,in BHDE model, the GSL is valid only for a much less value of $\Delta.$
 The dynamical system behavior of the present model predicts an asymptotically stable de Sitter epoch. This stability of the end phase is a ratification of the thermal evolution of the model, in which the end stage corresponds to a state of maximum entropy. Finally, we extended our work to perform the information criteria analysis to measure the significance of the present model in comparison with the standard $\Lambda$CDM. Our analysis shows that, apart from the interacting BHDE model, however, has significantly less evidence against $\Lambda$CDM in the AIC analysis. Meantime, the model essentially lacks support from BIC analysis.
 
To summarize, treating Barrow holographic dark energy as a dynamical vacuum with $\omega_{_\Lambda}=-1$ has some advantages over the models in which it is treated as dark energy of varying equation of state. The foremost thing is that, at non-interaction, the model mimics a $\Lambda$CDM behavior with an effective cosmological constant. Hence the model satisfies the GSL, irrespective of the value of the Barrow exponent, which is results in agreement to some previous works related to BHDE. The much less value estimated for the Barrow exponent $\Delta$ shows that the new area-entropy relation proposed by Barrow is very close to the Bekenstein-Hawking relation for horizon entropy , and it generally respect GSL. Even though the model predicts reasonably good background evolution, the AIC analysis shows less evidence in favor of the present model. In contrast, BIC analysis shows no evidence at all. However, it needs further detailed study regarding the evidence of this model against the standard one, which we will reserve for future work. 
 
\section*{Acknowledgments}
We are thankful to the reviewers for the valuable and insightful comments which helped us to improve the manuscript. We are thankful to Hassan Basari V T, Sarath N and Dr. Jerin Mohan N D for their valuable suggestions in data analysis. Nandhida Krishnan. P is thankful to  CUSAT and Govt. of Kerala for financial assistance.

\end{document}